\documentclass[conference]{IEEEtran}
\IEEEoverridecommandlockouts
\usepackage[letterpaper, left=0.75in, right=0.75in, bottom=0.75in, top=.75in]{geometry}
\usepackage{cite}
\usepackage{amsmath,amssymb,amsfonts}
\usepackage{todonotes}
\usepackage{multirow}
\usepackage{algorithmic}
\usepackage{graphicx}
\usepackage{textcomp}
\usepackage{xcolor}
\usepackage{nicematrix}
\usepackage{caption}
\usepackage{subcaption}
\usepackage{hyperref}
\def\BibTeX{{\rm B\kern-.05em{\sc i\kern-.025em b}\kern-.08em
    T\kern-.1667em\lower.7ex\hbox{E}\kern-.125emX}}

\newcommand{\ourmethod}{PRReach}

\newcommand{\ourmethodoff}{PRR-offline}
\newcommand{\ourmethodon}{PRR-online}

\newcommand{\Vol}{\text{Vol}}
\newcommand{\rtg}{\text{rtg}}

\makeatletter
\newcommand{\linebreakand}{%
  \end{@IEEEauthorhalign}
  \hfill\mbox{}\par
  \mbox{}\hfill\begin{@IEEEauthorhalign}
}
\makeatother

\title{\vspace*{.35cm}
PRREACH: Probabilistic Risk Assessment Using Reachability for UAV Control
\thanks{This work was partially supported by the FAA ASSURE Center of Excellence under projects A51 and A58, NSF Award 2118179, and the \href{https://www.faa.gov/newsroom/faa-selects-university-team-finalists-2024-data-challenge}{2024 FAA Data Challenge}.}
\thanks{$^{1}$School of Electrical Engineering and Computer Science, Oregon State University, Corvallis, OR, USA {\tt\small \{frondan,abbasho\}@oregonstate.edu}}
\thanks{$^{2}$Department of Electrical Engineering and Computer Science, Philadelphia, PA, USA {\tt\small \{hn384,sa3838,steven.weber\}@drexel.edu}}}

\author{Nicole Fronda$^{1}$, Hariharan Narayanan$^{2}$, Sadia Afrin Ananna$^{2}$, Steven Weber$^{2}$, Houssam Abbas$^{1}$}

\begin{document}

\maketitle

\begin{abstract}
We present a new approach for designing risk-bounded controllers for Uncrewed Aerial Vehicles (UAVs). 
Existing frameworks for assessing risk of UAV operations rely on knowing the conditional probability of an incident occurring given different causes. Limited data for computing these probabilities makes real-world implementation of these frameworks difficult.
Furthermore, existing frameworks do not include control methods for risk mitigation. Our approach relies on UAV dynamics, and employs reachability analysis for a probabilistic risk assessment over all feasible UAV trajectories. We use this holistic risk assessment to formulate a control optimization problem that minimally changes a UAV's existing control law to be bounded by an accepted risk threshold. We call our approach \ourmethod. Public and readily available UAV dynamics models and open source spatial data for mapping hazard outcomes enables practical implementation of \ourmethod~for both offline pre-flight and online in-flight risk assessment and mitigation. We evaluate \ourmethod~through simulation experiments on real-world data. Results show that \ourmethod~controllers reduce risk by up to 24\% offline, and up to 53\% online from classical controllers.
\end{abstract}


\section{Introduction} \label{sec:intro}

\begin{figure*}[ht]
  \centering
\includegraphics[width=.49\linewidth]{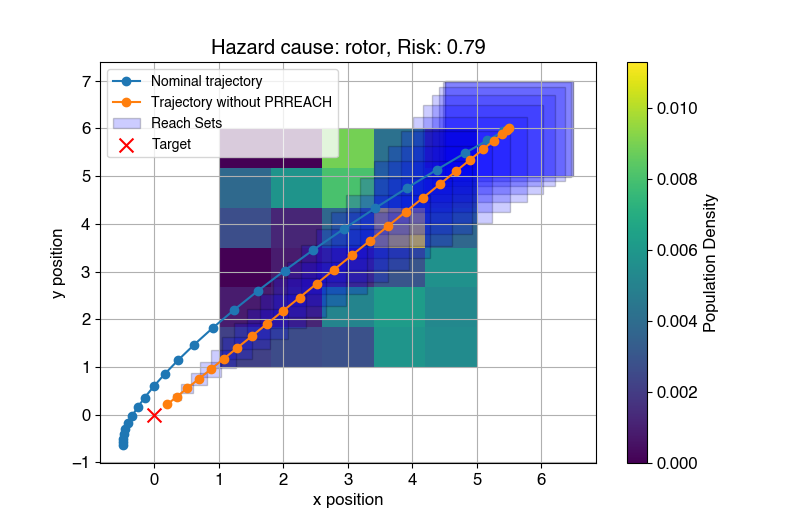} \hfill \includegraphics[width=.49\linewidth]{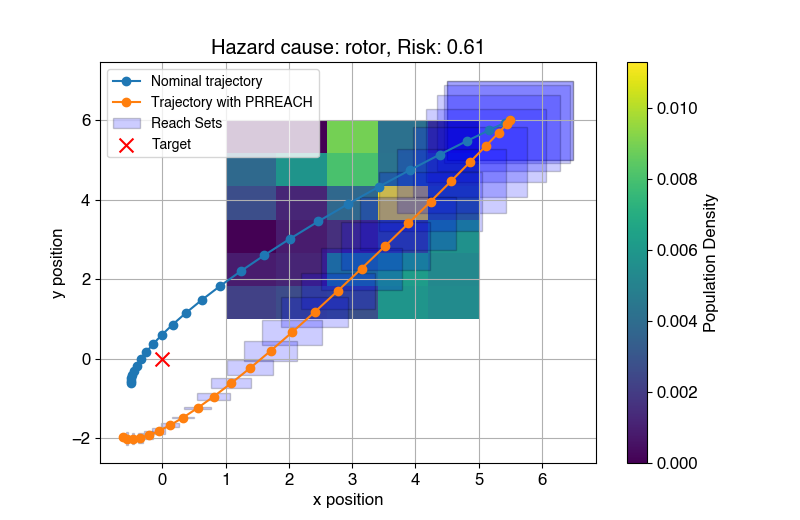}
    \caption{
    \small{On the left we see example UAV trajectories over a heatmap representing population density of city blocks in Philadelphia. The blue line shows the nominal trajectory over this area towards the target given by a red dot. The orange line shows the trajectory when the UAV experiences a deficient rotor, causing it to fly over a more densely populated area. The light blue rectangles show the reach sets of the UAV with the deficient rotor. 
    On the right we see the same heatmap and nominal UAV trajectory. The orange line here shows the UAV trajectory using a \ourmethod~controller, which directs the UAV away from the more densely populated regions. \ourmethod~used the reach sets to evaluate risk over all feasible trajectories of the UAV, and minimally change the UAV controller to maintain a bounded level of risk. This results in the reduced risk evaluation from 0.79 to 0.61 between the left and right plots, at the cost of the UAV on the right being  further from the target at the end of the time horizon. Best viewed in color.}}
    \label{fig: reachsetviz}
\end{figure*}

Uncrewed Aerial Vehicles (UAVs) have revolutionized various industries, from logistics and surveillance to environmental monitoring and emergency response. These operations often require a waiver \cite{Part_107_Waivers_2024} that provides a measure of risk of all potential \textit{hazard outcomes} - an undesirable event such as collision with a person or building. A common measure of risk is the probability of such an event occurring times its severity. Current risk management frameworks for UAVs use Probabilistic Risk Assessments (PRA) to calculate risk using conditional probabilities of a hazard outcome given a \textit{hazard cause} - a broken rotor, strong winds, sensor malfunctions, etc. However, these frameworks are severely constrained by lack of requisite data for computing these conditional probabilities, and are limited in their assessments and decisions. We propose a new approach for risk assessment and mitigation which, through reachability analysis, relates the dynamics of a UAV to the likelihood of hazard outcome as modeled by available spatial data of the UAV's flight environment. We call our approach \textbf{\ourmethod}, for \textbf{P}robabilistic \textbf{R}isk Assessment Using \textbf{Reach}ability.

\ourmethod~can be applied by regulators evaluating risk of proposed UAV operations, and can aid operators in developing risk mitigation measures. The process for operators would be as follows: 1) If a hazard cause occurs before flight, use a precomputed \ourmethod~controller to guarantee risk below a set threshold. Otherwise proceed with nominal flight controllers. 2) If no hazard cause is present initially but occurs later during flight either a) switch to the \ourmethod~controller that was precomputed \textit{offline} for the hazard cause, OR b) if time resources are available, compute a \ourmethod~controller \textit{online}. Figure \ref{fig: reachsetviz} shows example trajectories from \ourmethod.

\textbf{Related Work}
Existing PRA frameworks for traditional aircraft follow the procedural requirements for executing safety risk management within the FAA \cite{FAAOrder8040}. Proposed PRA frameworks for UAVs follow these same requirements, including the framework most recently presented in FAA Assure Report 21 \cite{A21_report}. In addition to relying on limited historical UAV data for modeling risk, a limitation which \ourmethod~does not have, this approach only considers categorical hazard causes - does wind exist or not. \ourmethod~can incorporate continuous hazard causes, such as wind speed into its risk assessment. The approach in \cite{A21_report} is also a purely offline framework, meant to give recommendations for whether or not an operation can take flight. It does not provide a method for in-flight control. In contrast, \ourmethod~allows for offline and online risk assessment, and produces risk-bounded controllers. Furthermore, \cite{A21_report} averages risk over all areas in the environment, and does not consider where the UAV can actually go according to its dynamics. Instead, \ourmethod~uses reachability analysis to compute risk over all feasible trajectories for a risk evaluation.

Reachability analysis is the computation of the set of future states of a dynamical system from some initial set of states. These sets are called \textit{reach sets}. It has been widely used as a tool in motion planning and control for many different types of dynamical systems. Algorithms are available for computing the reach sets of discrete-time and continuous-time linear, non-linear, deterministic and stochastic systems, with many specializations for various cases. See \cite{althoff21reachreview} and \cite{bak17hylaa} and the references therein for a background on the most common ways of computing reach sets and associated software tools. In this work, we use zonotopes to represent reach sets as they are closed under linear transformations. Conversion from zonotopes to other set representations is possible when necessary. For example, \cite{zono_to_ellipsoid} describes scalable conversions between zontopes and ellipsoids.

Existing control methods for robotic systems which use reachability include the works \cite{kousik2019technicalreportsafeaggressive, radius_2023,michaux2024redefined}. These works are similar to ours in that they use reach sets within an optimization problem to ultimately find trajectories that are bounded by some risk evaluation. However, \ourmethod~differs in that it optimizes the generation of the reach sets themselves, and not over trajectories within the reach set. \ourmethod~yields a risk-aware \textit{controller} that can produce multiple trajectories for a UAV given different initial states, each guaranteed to not exceed some level of risk. As far as we know, this approach is the first that optimizes a controller in this way.

\noindent \textbf{Contributions}
In this paper we present:
\begin{enumerate}
    \item A PRA method for computing risk over reach sets
    \item A formulation for a risk-bounded control optimization
    \item A solution to the optimization called \ourmethod, which we evaluate in simulation using real-world data and compare with a classical controller.
\end{enumerate}



\section{Preliminaries}

We define a UAV CONOPS (Concept of Operations) to include the set of possible initial states of the UAV, the destination of the UAV, the time horizon over which the UAV must travel from its initial state to the destination, and the \textit{generalized dynamics} of the UAV.

For a set of hazard causes $C$, the generalized dynamics of the UAV is denoted as $F \triangleq (f_c \mid c \in \mathcal{C})$. This is a collection of \emph{linear, discrete time} dynamics under each possible hazard cause $c$, including when no hazard cause occurs. That is, given the occurrence of some hazard cause $c$, the UAV follows the dynamics:
\begin{align}
    \textbf{x}[k+1] &= f_c(\mathbf{x}[k],\mathbf{u}[k],\mathbf{w}[k])
\end{align}

where $\mathbf{x}[k] \in \mathcal{X}$ is the UAV state at time index $k$, $\mathbf{u}[k]$ is the control input to the UAV at time index $k$, and $\mathbf{w}[k]$ represents an external disturbance to the UAV, such as wind, at time index $k$. Because we focus on the linear case, we can write the dynamics function $f_c$ as:
\begin{align}
    f_c(\mathbf{x}[k],\mathbf{u}[k],\mathbf{w}[k]) = \mathbf{A}_c \mathbf{x}[k] + \mathbf{B}_c \mathbf{u}[k] + \mathbf{B'}_c \mathbf{w}[k]
\end{align}

where $\mathbf{A}_c$ and $\mathbf{B}_c$ are the drift and control matrices associated with hazard cause $c$, and $\mathbf{B'}_c$ relates the external disturbances associated with $c$ to the UAV's state.

Using these matrices for the dynamics, we can obtain a feedback controller with gain matrix $\mathbf{K}_c$ by solving the Linear Quadratic Regular (LQR) equations \cite{underactuated}. Under this controller, the UAV state at any time $k > 0$ is:

\begin{align}
    \mathbf{x}[k] = (\mathbf{A}_c - \mathbf{B}_c \mathbf{K}_c)^k \mathbf{x}[0] + \mathbf{B'}_c \mathbf{w}[k]
\end{align}


Given a \textit{set} of possible initial states of the UAV, the set of all feasible states - the \textit{reach set} -  of the UAV at any time $k > 0$ can be computed using reachability analysis. 
Let us represent the set of all possible initial states of a UAV $X_0$, and the set of all possible disturbances $W$. The reach set at time $k$, denoted by $R_c(k)$, is then:
\begin{align}
    R_c(k) = (\mathbf{A}_c - \mathbf{B}_c \mathbf{K}_c)^k X_0 + \mathbf{B'}_c W
\end{align}



Following \cite{althoff21reachreview}, we represent a reach set as a \textit{zonotope}, a type of geometric set representation that is closed under linear operations. A zonotope $Z$ is defined as 
$Z = \{ c + \mathbf{G} \beta_i ~|~ \beta_i \in [-1, 1] \}$
with a center $c$ and $\mathbf{G}$ a matrix of generator vectors. We refer to \cite{althoff21reachreview} for more details of reach set computation with zonotopes.


\section{Mathematical Formulation of \ourmethod}
\subsection{Risk over a Reach Set}
For every hazard outcome $\theta \in \Theta$, for $\Theta$ being the set of considered hazard outcomes, we define a hazard map $\lambda_{\theta} : \mathcal{X} \longrightarrow \mathbb{R}$ that relates a UAV state, which includes its position, with the risk of the hazard outcome. In this work, we consider outcomes of uniform severity. For example, if $\theta$ is collision with a person, $\lambda_{\theta}(\mathbf{x})$ represents the probability of the UAV colliding with a person at the position in $\mathbf{x}$, which has the same severity regardless of person or position.

Let $p_{\theta}(R_c(k))$ be the probability of a hazard outcome $\theta$ occurring given a reach set as the total probability over all states in the reach set, divided by the volume of the set:

\begin{align} \label{eq: prob_hazard_rc}
    p_{\theta}(R_c(k)) \triangleq \frac{\int_{R_c(k)} \lambda_{\theta}(\mathbf{x}) \mathrm{d} \mathbf{x}}{\int_{R_c(k)} \mathrm{d} \mathbf{x}}, ~ \theta \in \Theta.
\end{align}


We can consider each reach set of a UAV CONOPS as a state in a Markov process where the transition probability from $R_c(k)$ to $R_c(k+1)$ is the probability that a hazard outcome does \textit{not} occur at $R_c(k)$: 

\begin{align}
    \mathbb{P}(R_c(k+1) \mid R_c(k)) = 1 - \displaystyle\sum_{\theta \in \Theta}p_{\theta}(R_c(k))
\end{align}


\noindent The probability of a UAV CONOPS being successful under a hazard cause $c$ and not experiencing a hazard outcome is:

\begin{align} \label{eq: prob_CONOPS_success}
    \rho_c = \displaystyle\prod_{k = 1}^T (1-\displaystyle\sum_{\theta \in \Theta}p_{\theta}(R_c(k)) )
\end{align}

\subsection{Problem Formulation}

We formalize computing a risk-bounded UAV controller as the following control optimization problem:

\begin{align} \label{eq: opt_formalization}
&\min_{\mathbf{K}_c'} ~ \|\mathbf{K}_c - \mathbf{K}_c'\|_F \\
&\text{s.t. } 1 - (\prod_{k = 1}^T 1- p_{\theta}(R_c(k))) \leq r \nonumber
\end{align}

\noindent where $\|\ldots\|_F$ is the Frobenius norm and $r$ is the risk threshold set by the operator. Essentially, given a hazard cause $c$ and probability function $p_{\theta}$ for a hazard outcome, solving the optimization problem will yield a new feedback controller gain matrix $\mathbf{K}_c'$ that can be used to control the UAV under dynamics $f_c$ associated with the hazard cause, that is close to the original controller, but is also \emph{guaranteed to not exceed the desired risk level}.


\subsection{Solution}

We first consider the special case of time horizon $T=1$, with initial set $X_0$, a hazard cause $c$ and the associated dynamics, and a set $H \subset \mathcal{X}$ which represents the area where the hazard outcome can occur (for example, a densely populated area where the UAV may collide with a person). Outside $H$, $\lambda(\mathbf{x}) = 0$. In this special case we need only compute risk over the first reach set, $R_c(1)$. The constraint in \eqref{eq: opt_formalization} then reduces to $p_{\theta}(R_c(1)) \leq r$, where $R_c(1) = (\mathbf{A}_c - \mathbf{B}_c \mathbf{K_c}) X_0 + \mathbf{B'}_c W$. The zonotope $W$ represents the set of external disturbances associated with hazard cause $c$. Substituting this into \eqref{eq: opt_formalization}:

\begin{subequations}
\begin{equation}
\min_{\mathbf{K}_c'} ~ \|\mathbf{K}_c - \mathbf{K}_c'\|_F \\
\end{equation}
\begin{equation} \label{eq: one_step_opt_constraint}
\text{s.t. } \frac{\int_{((\mathbf{A}_c - \mathbf{B}_c \mathbf{K}_c') X_0 + \mathbf{B'}_c W) \cap H} \lambda_{\theta}(\mathbf{x}) \mathrm{d} \mathbf{x}}{\int_{(\mathbf{A}_c - \mathbf{B}_c\mathbf{K}_c')X_0 + \mathbf{B'}_c W} \mathrm{d} \mathbf{x}} \leq r
\end{equation} 
\end{subequations}

\noindent We then substitute $\mathbf{D} = \mathbf{A}_c - \mathbf{B}_c\mathbf{K}_c'$ and minimize over $\mathbf{D}$

\begin{subequations}
\begin{equation}
\min_{\mathbf{D}} ~ \|(\mathbf{A}_c - \mathbf{B}_c\mathbf{K}_c) - \mathbf{D}\|_F \\
\end{equation}
\begin{equation} \label{eq: one_step_opt_constraint_D}
\text{s.t. } \frac{\int_{(\mathbf{D} X_0 + \mathbf{B'}_c W) \cap H} \lambda_{\theta}(\mathbf{x}) \mathrm{d} \mathbf{x}}{\int_{\mathbf{D} X_0 + \mathbf{B'}_c W} \mathrm{d} \mathbf{x}} \leq r
\end{equation} 
\end{subequations}

Note that the denominator of the LHS in \eqref{eq: one_step_opt_constraint_D} is simply the \textit{volume} of the zonotope $\mathbf{D} X_0 + \mathbf{B'}_c W$, which we refer to with operator $\Vol(\ldots)$. See \cite{mcmullen_vol_zonotope} for the computation of zonotope volumes.
We substitute the denominator in \eqref{eq: one_step_opt_constraint_D} with $\Vol(\mathbf{D} X_0 + \mathbf{B'}_c W)$.

We rewrite the integral in the numerator of the LHS in \eqref{eq: one_step_opt_constraint_D} to 

\begin{align} \label{eq: prob_hazard_num}
    \int_{(\mathbf{D} X_0 + \mathbf{B'}_c W) \cap H} \lambda_{\theta}(\mathbf{x}) \mathrm{d} \mathbf{x} = \int_{\mathbf{D} X_0 + \mathbf{B'}_c W} \lambda_{\theta}(\mathbf{x}) \cdot \mathbf {1}_{H} (\mathbf{x}) \mathrm{d} \mathbf{x}
\end{align}

\noindent where $\mathbf {1}_{H}(\mathbf{x})$ is an indicator function that is 1 if $\mathbf{x}$ is in $H$ and 0 otherwise. 


We decompose the zonotope $\mathbf{D} X_0 + \mathbf{B'}_c W$ into $\mathcal{S}$ simplices, where the $i$th simplex is given by $S_i$. We can now rewrite \eqref{eq: prob_hazard_num} as:

\begin{align} \label{eq: prob_hazard_num_decom}
    \int_{\mathbf{D} X_0 + \mathbf{B'}_c W} \lambda_{\theta}(\mathbf{x}) \cdot \mathbf {1}_{H} (\mathbf{x}) \mathrm{d} \mathbf{x} = \sum_{i=1}^\mathcal{S} \int_{S_i} \lambda_{\theta}(\mathbf{x}) \cdot \mathbf {1}_{H} (\mathbf{x}) \mathrm{d} \mathbf{x}
\end{align}



Recall that $\lambda_{\theta}$ is a hazard map. To obtain an analytically tractable optimization, we fit an $M$-degree polynomial approximation of $\lambda_{\theta}(\mathbf{x}) \cdot \mathbf{1}_{H}$, $P_{\theta,H}(\mathbf{x}) \approx \lambda_{\theta}(\mathbf{x}) \cdot \mathbf{1}_{H}(x)$. We write this polynomial as a sum of powers of linear forms $\mathbf{l}_m$:

\begin{equation}
    \lambda_{\theta}(\mathbf{x}) \cdot \mathbf{1}_{H}(\mathbf{x}) \approx P_{\theta,H} = \sum_{m=1}^M (\mathbf{l}_m^T\mathbf{x})^m.
\end{equation}

Approximating the hazard map as a polynomial allows us to rewrite each integral in the summand in \eqref{eq: prob_hazard_num_decom} using Theorem 1 from \cite{loera} to integrate polynomials over simplices using:

\begin{align} \label{eq: prob_hazard_rc_expanded}
& \int_{S_i} \lambda_{\theta}(\mathbf{x}) \cdot \mathbf {1}_{H} (\mathbf{x}) \mathrm{d} \mathbf{x} \approx \sum_{m=1}^M 2 \Vol(S_i) \frac{m!}{(m+2)!} \times \\
& \sum_{j \in \mathbb{N}^{d+1}, |j|=m} (\mathbf{l}_m^T S_{i,1})^{j_1} \ldots (\mathbf{l}_m^T S_{i,d+1})^{j_{d+1}} \nonumber
\end{align}

\noindent where $|j|=\sum_{u=1}^{d+1} j_u$, and $S_{i, j}$ denotes the $j$th vertex of simplex $S_i$ of dimension $d$. We compute the volume of a simplex $\Vol(S_i)$ as shown in \cite{simplex_vol}.

Substituting all this back into the numerator of the LHS in \eqref{eq: one_step_opt_constraint_D} we have

\begin{subequations}
\begin{equation}
\min_{\mathbf{D}} ~ \|(\mathbf{A}_c - \mathbf{B}_c\mathbf{K}_c) - \mathbf{D}\|_F \\
\end{equation}
\begin{equation}
\text{s.t. } ~ \frac{\eta}{\Vol(\mathbf{D} X_0 + \mathbf{B'}_c W)} \leq r 
\end{equation}
\begin{align}
& \eta = \sum_{i=1}^\mathcal{S} \sum_{m=1}^M 2 \Vol( S_i) \frac{m!}{(m+2)!} \times [ \nonumber \\ 
& \sum_{j \in \mathbb{N}^{d+1}, |j|=m} (\mathbf{l}_m^T S_{i,1})^{j_1} \ldots (\mathbf{l}_m^T S_{i,d+1})^{j_{d+1}} ]
\end{align}
\end{subequations}

The above result is for the case where $T=1$. We generalize this to multiple time steps $1 \leq k \leq T$ of the CONOPS:

\begin{subequations} \label{eq: optimization_final}
\begin{equation}
\min_{\mathbf{D}} ~ \|(\mathbf{A}_c - \mathbf{B}_c\mathbf{K}_c) - \mathbf{D}\|_F
\end{equation}
\begin{equation} \label{eq:opt_final_risk_thresh}
\text{s.t. } \frac{\eta_k}{\Vol(Z^{(k)})} \leq r_k 
\end{equation}
\begin{align}
& \eta_k = \sum_{i=1}^{\mathcal{S}^{(k)}} \sum_{m=1}^M 2 \Vol( S_{k,i}) \frac{m!}{(m+2)!} \times [ \nonumber \\ 
& \sum_{j \in \mathbb{N}^{d_1}, |j|=m} (\mathbf{l}_m^T S^{(k)}_{i,1})^{j_1} \ldots (\mathbf{l}_m^T S^{(k)}_{i,d+1})^{j_{d+1}} ]
\end{align}
\begin{equation}
Z^{(k)} = \mathbf{D}^k X_0 + \sum^{k-1}_{l=1} \mathbf{D}^{l}\mathbf{B'}_c W
\end{equation}
\end{subequations}

\noindent where zonotope $Z^{(k)}$ is decomposed into $\mathcal{S}^{(k)}$ number of simplices, with each $i$th simplex indexed as $S^{(k)}_{i}$. In \eqref{eq:opt_final_risk_thresh}, $r_k$ is the desired risk threshold for time $k$, and set such that $1 - \prod_k^T (1 - r_k) \leq r$.

We now have a purely analytical expression of our problem, with computable gradients for the objective and constraints. Taking as input the dynamics matrices for a hazard cause, a zonotope set of initial states and its simplicial decomposition, and a polynomial probability function written as the sum of powers of linear forms, Problem \eqref{eq: optimization_final} can be given to a general global optimizer to be solved. Once an optimal $\mathbf{D}$ is found after solving, we obtain the optimal gain matrix $\mathbf{K}_c'$ through
\begin{equation} \label{eq: get_K_opt}
\mathbf{K}_c' = \mathbf{B}_c^{\dagger} (\mathbf{A}_c - \mathbf{D})
\end{equation}
where $\dagger$ indicates the Moore-Penrose pseudoinverse which can be found using matrix singular value decomposition.

\section{Implementation and Experiments}
\subsection{Hazard Outcome Maps}
We consider two hazard outcomes maps: $\lambda_{\theta1}$ to represent probability of collision with a person and $\lambda_{\theta2}$ to represent probability of collision with a building. Both $\lambda_{\theta1}$ and $\lambda_{\theta2}$ are 3-degree polynomials approximating the population and building density of city blocks in Philadelphia, PA, available through the open data repository OpenDataPhilly \cite{OpenDataPhilly}. Figures \ref{fig: reachsetviz} and \ref{fig: online_v_offline} show examples with  $\lambda_{\theta1}$ and $\lambda_{\theta2}$ respectively.

\subsection{UAV Dynamics}
For our experiment simulations, we use the following 12 dimensional state representation for a quadrotor UAV from \cite{sabatino2015quadrotor}: $\mathbf{x} = [x, y, z, \phi, \gamma, \psi, u,v,w, p, q, r]^T$ where $x,y,z$ represent its 3D position, $\phi, \gamma, \psi$ represents roll, pitch, and yaw, $u,v,w$ represent the linear velocities, and $p, q, r$ represent the roll rate, pitch rate, and yaw rate. We use the following to represent the control input of the UAV: $\mathbf{u} = [f_t, \tau_x, \tau_y, \tau_z]^T$ where $f_t$ is the total thrust, and $\tau_x, \tau_y, \tau_z$ are the torques controlling roll, pitch, and yaw.  In practice, ore accurate dynamics can be used so long as they are linear. 

In all of the following, let $g$ represent the gravity, $m$ the mass of the UAV, and $J_x, J_y, J_z$ the moments of inertia around the $x,y,z$ axes of the UAV, respectively. Under no hazard cause, the UAV dynamics matrices take the form:

\setcounter{MaxMatrixCols}{20}
\begin{align}
\mathbf{A} = \begin{bNiceArray}{cccccccccc}[small, columns-width = 3mm ]
\Block{3-3}{\mathbf{0}} & & \Block{3-3}{\mathbf{I}} & & \Block{3-3}{\mathbf{0}} & & \Block{3-3}{\mathbf{0}} & & \\ 
& & & & & & & & \\
& & & & & & & & \\
\Block{3-3}{\mathbf{0}} & & \Block{3-3}{\mathbf{0}} & & \Block{3-3}{\mathbf{0}} & & \Block{3-3}{\mathbf{0}} & & \\ 
& & & & & & & & \\
& & & & & & & & \\
0 & -g & 0 \Block{3-3}{\mathbf{0}} & & \Block{3-3}{\mathbf{0}} & & \Block{3-3}{\mathbf{0}} & & \\ 
g & 0 & 0 & & & & & & \\
0 & 0 & 0 & & & & & & \\
\Block{3-3}{\mathbf{0}} & & \Block{3-3}{\mathbf{0}} & & \Block{3-3}{\mathbf{I}} & & \Block{3-3}{\mathbf{0}} & & \\ 
& & & & & & & & \\
& &  & & & & & & \\
\end{bNiceArray},
\mathbf{B} = \begin{bNiceArray}{cccc}[small, columns-width = 3mm ]
\Block{3-4}{\mathbf{0}} & & & \\
& & & \\
& & & \\
0 & 1/J_x & 0 & 0 \\
0 & 0 & 1/J_y & 0 \\
0 & 0 & 0 & 1/J_z \\
0 & 0 & 0 & 0 \\
0 & 0 & 0 & 0 \\
1/m & 0 & 0 & 0 \\
\Block{3-4}{\mathbf{0}} & & & \\
& & & \\
& & & \\
\end{bNiceArray}
\end{align}



We also simulate the following hazard causes for our experiments.

\begin{enumerate}
    \item \textbf{Deficient Rotor}: Reduces overall control of the UAV, which we model by scaling the control matrix by a coefficient $\alpha_{def}$. The dynamics matrices become:
    \begin{align}
        \mathbf{A}_{dr} = \mathbf{A},~ \mathbf{B}_{dr} = \alpha_{dr}\mathbf{B}
    \end{align}
    \item \textbf{Sensor Error}: Reduces accuracy in controlling the UAV's position. We model this by altering how the UAV's current position affects its next position. The further the UAV is from its origin the further away it will drift.
        \begin{align}
        \mathbf{A}_{se} &= \begin{bNiceArray}{cccccccccc}[small, columns-width = 3mm ]
        \alpha^{x}_{se} & 0 & 0 \Block{3-3}{\mathbf{I}} & & \Block{3-3}{\mathbf{0}} & & \Block{3-3}{\mathbf{0}} & & \\ 
        0 & \alpha^{y}_{se} & 0 & & & & & & \\
        0 & 0 & 0 & & & & & & \\
        \Block{3-3}{\mathbf{0}} & & \Block{3-3}{\mathbf{0}} & & \Block{3-3}{\mathbf{0}} & & \Block{3-3}{\mathbf{0}} & & \\ 
        & & & & & & & & \\
        & & & & & & & & \\
        0 & -g & 0 \Block{3-3}{\mathbf{0}} & & \Block{3-3}{\mathbf{0}} & & \Block{3-3}{\mathbf{0}} & & \\ 
        g & 0 & 0 & & & & & & \\
        0 & 0 & 0 & & & & & & \\
        \Block{3-3}{\mathbf{0}} & & \Block{3-3}{\mathbf{0}} & & \Block{3-3}{\mathbf{I}} & & \Block{3-3}{\mathbf{0}} & & \\ 
        & & & & & & & & \\
        & &  & & & & & & \\
        \end{bNiceArray},~ 
        \mathbf{B}_{se} = \mathbf{B},
        \end{align}
    \item \textbf{Wind Disturbance} Pushes the UAV outside of its nominal trajectory. We model the wind disturbance as  $\mathbf{w}[k] = [f_x[k], f_y[k], f_z[k], \tau_{x}[k], \tau_{y}[k], \tau_{z}[k]]^T$, with $f_x[k], f_y[k], f_z[k]$ being the directional forces and $\tau_{x}[k], \tau_{y}[k], \tau_{z}[k]$ the torques on the UAV from the wind at time $k$.
    The effect of $\mathbf{w}[k]$ on the UAV's state added by the disturbance matrix
    \begin{align}
        \mathbf{B'}_{wd} = \begin{bNiceArray}{cccccc}[small, columns-width = 3mm]
        \Block{3-3}{\mathbf{0}} & & & \Block{3-3}{\mathbf{0}} & &\\
        & & & & & \\
        & & & & & \\
        \Block{3-3}{\mathbf{0}} & & & 1/J_x & 0 & 0 \\
        & & & 0 & 1/J_y & 0 \\
        & & & 0 & 0 & 1/J_z \\
        1/m & 0 & 0 & \Block{3-3}{\mathbf{0}} & & \\
        0 & 1/m & 0 & & & \\
        0 & 0 & 1/m & & & \\
        \Block{3-3}{\mathbf{0}} & & & \Block{3-3}{\mathbf{0}} & &\\
        & & & & & \\
        & & & & &
        \end{bNiceArray}
    \end{align}
    Let zonotope $W$ encompass the range of all $\mathbf{w}[k]$ generated by distribution $\mathcal{N}([0.05, 0.31, 0, -0.005, -0.03]^T, 0.03)$, units $m/s$.
\end{enumerate}

The above matrices describe continuous-time dynamics which we convert to discrete-time using sampling time $d_t = 0.01s$. We set parameters $m=1.5kg$, $g=9.81m/s^2$, $J_x = J_y = 0.02kg\cdot m^2$, $J_z=0.04kg\cdot m^2$, $\alpha_{dr} = 0.4$, $\alpha^x_{se} = \alpha^y_{se}=0.6$ and $T=25$. Under each of the hazard causes, we compute an LQR controller as a baseline controller not optimized to reduce risk of any hazard outcome.

\subsection{Experimental Setup}
We implement the operational process for \ourmethod~described in Section \ref{sec:intro}, and run two experiments. The first precomputes \ourmethod~controllers \textit{offline} assuming the presence of a hazard cause before flight. The second computes \ourmethod~controllers \textit{online} after a hazard cause occurs.

We set up a CONOPS in which the UAV must maintain its altitude and fly to the origin from an initial position selected from within a 2D square zonotope centered at $(5.5, 6)$. We set each $r_k$ in \eqref{eq: optimization_final} to the risk over the reach set at time $k$ generated by the LQR controller under the no-hazard dynamics. This ensures the resulting controllers do not produce trajectories whose overall risk exceeds that of trajectories produced under normal conditions.

For the first experiment, we solve \eqref{eq: optimization_final} to pre-compute a risk bounded controller for each hazard outcome map and hazard cause combination. 
We call the result a \ourmethodoff~controller. We compare the trajectory it generates from the initial position with the trajectory generated by the LQR controller in terms of risk and final distance to the target.

For the second experiment, we solve \eqref{eq: optimization_final} online for 100 simulated flights, each from different initial positions within the zonotope, and with a hazard cause occurring at a randomly selected time $k'$. This differs from the first experiment in that \eqref{eq: optimization_final} is solved for only the remaining time steps $k'\ldots T$. We call this result a \ourmethodon~controller.
We compare the risk and final distance to target of the trajectories generated by \ourmethodoff~and \ourmethodon~controllers from time $k'$.



Simulations were implemented in Python using SciPy for solving the optimization. Experiments were run on a computer with an 8-core 3.2Ghz processor and access up to 32GB RAM. 


\section{Results}
For the following discussion, we use ``risk-to-go" (rtg) to refer to the aggregate risk of a hazard outcome $\theta$ over a time period [k, T] given the occurrence of a hazard cause $c$ at time $k$: $\rtg(\theta, c, k, \text{controller name}, \text{[k, T]})$.
For example, $\rtg(\theta_1, \text{wd}, 5, \text{LQR}, \text{[5,T]})$ refers to the rtg of collision with a person from time $k=5$ to the end of the CONOPS, given the wind hazard cause occurring at $k=5$.

\subsection{Reducing Risk for Greater Distance to Target}
From our first experiment, we found that a trajectory produced by the \ourmethodoff~controller has lower starting rtg than that of the LQR controller. In other words, $\rtg(\theta, c, 0, \text{\ourmethodoff}, [0, T]) < \rtg(\theta, c, 0, \text{LQR}, [0, T])$ for all hazard outcomes and causes we considered. The trade-off for lower risk compared to the LQR controller is a greater distance between the final state in the trajectory and the target destination. Table \ref{tab:offline_tradeoffs} shows this risk reduction and increase in distance of the \ourmethodoff~controller produced trajectories as a percentage of those of the LQR controller.

\begin{table}[t]
\centering
\resizebox{\linewidth}{!}{%
\begin{tabular}{|c|cc|cc|}
\hline
                      & \multicolumn{2}{c|}{\textbf{\% Reduction in Risk}} & \multicolumn{2}{c|}{\textbf{\% Increased Distance to Target}} \\ \cline{2-5} 
\textbf{Hazard Cause} & \textbf{Person}     & \textbf{Building}     & \textbf{Person}           & \textbf{Building}          \\ \hline
\textbf{Deficient Rotor} & 23.55\% & 1.62\% & 598.90\% & 780.10\% \\
\textbf{Sensor Error}    & 8.10\%  & 1.40\% & 33.87\%  & 89.50\%  \\
\textbf{Wind}            & 25.52\% & 3.27\% & 3.15\%   & 0.31\%   \\ \hline
\end{tabular}%
}
\caption{\small{Risk reduction and increase in final distance to target of the trajectories produced by the \ourmethodoff~controller as a percentage of the risk and final distance to target of trajectories produced by the LQR controller. The trajectories produced by the \ourmethodoff~controller reduce the risk of collision with a person or building in the presence of a hazard cause, but at the expense of greater distance to the target at the end of the time horizon, for all three hazard causes.}}
\label{tab:offline_tradeoffs}
\end{table}

\begin{table*}[t]
\small
\centering
\resizebox{\textwidth}{!}{%
\begin{tabular}{|c|cccc|cccc|}
\hline
 &
  \multicolumn{4}{c|}{\textbf{Average \% Reduction in Risk}} &
  \multicolumn{4}{c|}{\textbf{Average \% Increased Distance to Target}} \\ \cline{2-9} 
 &
  \multicolumn{2}{c|}{\textbf{Person}} &
  \multicolumn{2}{c|}{\textbf{Building}} &
  \multicolumn{2}{c|}{\textbf{Person}} &
  \multicolumn{2}{c|}{\textbf{Building}} \\ \cline{2-9} 
\textbf{Hazard Cause} &
  \multicolumn{1}{c|}{\textbf{\ourmethodoff}} &
  \multicolumn{1}{c|}{\textbf{\ourmethodon}} &
  \multicolumn{1}{c|}{\textbf{\ourmethodoff}} &
  \textbf{\ourmethodon} &
  \multicolumn{1}{c|}{\textbf{\ourmethodoff}} &
  \multicolumn{1}{c|}{\textbf{\ourmethodon}} &
  \multicolumn{1}{c|}{\textbf{\ourmethodoff}} &
  \textbf{\ourmethodon} \\ \hline
\textbf{Deficient Rotor} &
  23.62\% &
  \multicolumn{1}{c|}{47.84\%} &
  3.98\% &
  24.25\% &
  311.07\% &
  \multicolumn{1}{c|}{42.70\%} &
  \multicolumn{1}{c|}{417.17\%} &
  80.46\% \\
\textbf{Sensor Error} &
  8.46\% &
  \multicolumn{1}{c|}{43.90\%} &
  3.29\% &
  25.03\% &
  37.87\% &
  \multicolumn{1}{c|}{-27.58\%} &
  \multicolumn{1}{c|}{108.89\%} &
  -16.77\% \\
\textbf{Wind} &
  24.02\% &
  \multicolumn{1}{c|}{53.46\%} &
  3.18\% &
  29.39\% &
  -4.05\% &
  \multicolumn{1}{c|}{-27.83\%} &
  \multicolumn{1}{c|}{2.16\%} &
  -37.19\% \\ \hline
\end{tabular}%
}
\caption{\small{Average risk reduction and increase in final distance to target of the trajectories produced by the \ourmethodoff~and \ourmethodon~controllers as a percentage of the risk and final distance to target of trajectories produced by the LQR controller. Results were averaged over 100 trajectories in which the hazard cause occurred at random times in flight. Trajectories produced by \ourmethodon~controllers resulted in greater risk reduction than \ourmethodoff~controllers, as well as lower distance to the target. In some cases, \ourmethod~would produce trajectories with lower distance to the target compared to the LQR controller, as indicated by the negative values.}}
\label{tab:online_v_offline}
\end{table*}

\subsection{Lowering In-Flight Risk with Online Controllers}

\begin{figure}[]
  \centering
\includegraphics[width=\linewidth]{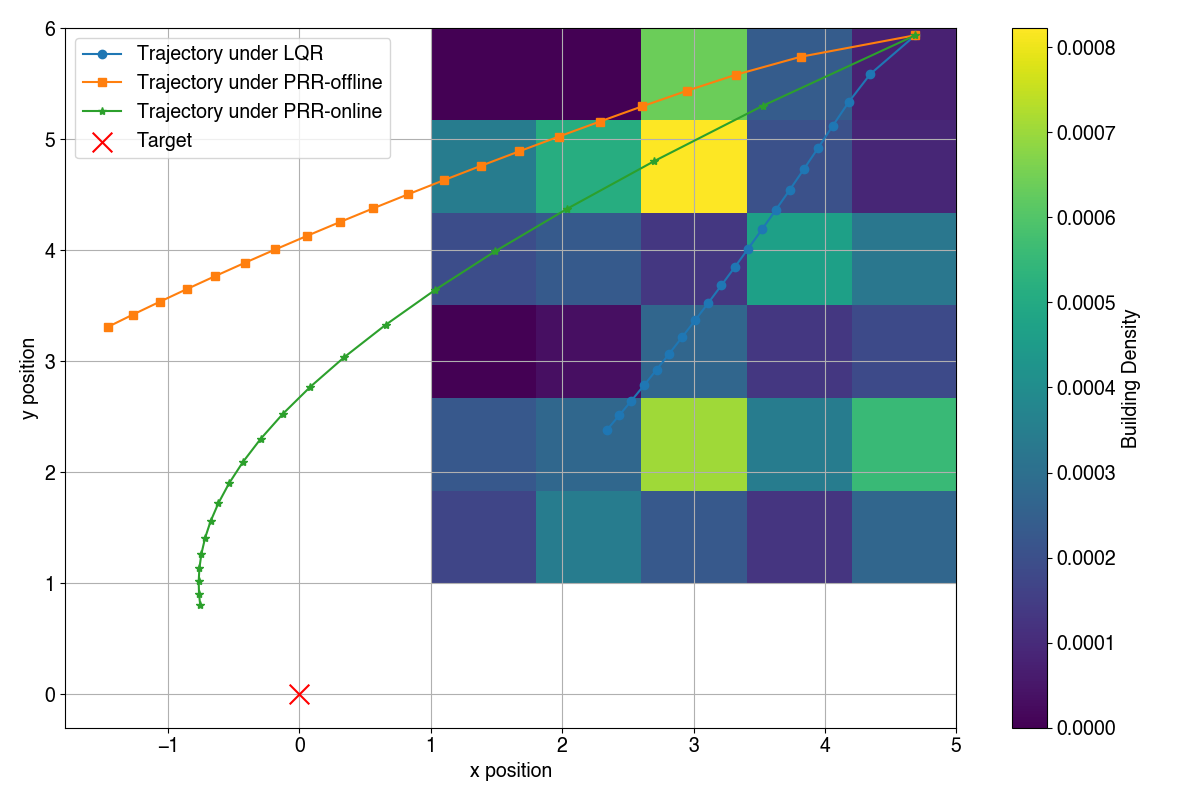}
    \caption{\small{Trajectories produced by an LQR controller (blue), a \ourmethodoff~controller (orange), and a \ourmethodon~controller (green) under the wind hazard. The hazard cause occurs at $k=5$, and the trajectories depict the UAV's simulated path under each of the controllers over $k=5\ldots T$. The heatmap shows the building density. Outside the heatmap there is no risk of collision with a building. The \ourmethod~controllers produce trajectories that spend less time over buildings as compared to the LQR controller, and the trajectory from the \ourmethodon~controller ends up closest to the target point (red X).}}
    \label{fig: online_v_offline}
\end{figure}

We found in our second experiment that by re-solving \eqref{eq: optimization_final} at the time the hazard cause occurs and using the resulting \ourmethodon~controller, rather than simply switching to the \ourmethodoff~controller, the resulting trajectory on average will have even lower risk. That is, the rtg from the time the hazard occurs can be further reduced by re-optimizing the UAV controller with up to date information at that time point. Specifically, $\rtg(\theta, c, k', \text{\ourmethodon}, [k', T]) < \rtg(\theta, c, k', \text{\ourmethodoff}, [k', T]) < \rtg(\theta, c, k', \text{LQR}, [k', T])$, where $k'$ is the time the hazard cause $c$ occurs.
Furthermore, the distance between the final state in the trajectory and the target destination was on average closer for trajectories produced from the \ourmethodon~than the \ourmethodoff~controller. For the sensor error and wind hazard causes, the trajectories would result in a final position that is closer to the target than that of the LQR controller. Table \ref{tab:online_v_offline} summarizes these results, and Figure \ref{fig: online_v_offline} visualizes trajectories from a single simulation.

\subsection{Optimization Runtime}

\begin{table}
\centering
\resizebox{\columnwidth}{!}{%
\begin{tabular}{|c|ccc|ccc|}
\hline
                & \multicolumn{3}{c|}{\textbf{Person}} & \multicolumn{3}{c|}{\textbf{Building}} \\ \cline{2-7} 
\textbf{Hazard Cause}    & \textbf{Average}  & \textbf{Max}      & \textbf{Std Dev.}  & \textbf{Average}  & \textbf{Max}     & \textbf{Std Dev.} \\ \hline
\textbf{Deficient Rotor} & 8.829    & 11.457   & \textbf{0.812}     & 9.524    & 11.070  & 1.147    \\
\textbf{Sensor Error}    & \textbf{7.994}    & \textbf{9.969}    & 1.031     & \textbf{9.424}    & \textbf{10.763}  & \textbf{0.965}    \\
\textbf{Wind}            & 53.630   & 396.593  & 68.337    & 16.899   & 38.678  & 10.774   \\ \hline
\end{tabular}%
}
\caption{The maximum, average, and standard deviation of runtime (in seconds) to compute a \ourmethod~controller. Runtimes depend on UAV dynamics, with the dynamics under the sensor error hazard cause having lowest runtime.}
\label{tab:runtime}
\end{table}

The time complexity for \ourmethod~depends on computation of the reach sets which in the worst case takes $\mathcal{O}(T(n^2 + n^2g_0 + n(g_0+g_w)))$ for time horizon $T$, state size of $n$, $g_0$ the number of generator vectors of zonotope $X_0$, and $g_w$ the number of generator vectors of $W$. In our experiments, $X_0$ and $W$ each have 2 generators, and we apply a heuristic generator reduction technique to keep each computed zonotope reach set at 2 generators.

Runtime to solve \eqref{eq: optimization_final} varies depending on the hazard cause dynamics, with the range of averages for the dynamics we chose being 8-53 seconds. Over all simulations, the maximum time to compute a controller took almost 400 seconds under the wind hazard cause. Thus, computing a controller online for the wind hazard cause may not be feasible.
The maximum, average, and standard deviation of runtimes 
in our second experiment are reported in Table \ref{tab:runtime}.

\subsection{\ourmethod~in Practice}
For a UAV CONOPS, if some deviation to the target can be tolerated, and thus accounted for with longer flight times, \ourmethod~provides an offline tool for operators to prepare controllers for mitigating risk of hazard outcomes under different hazard causes. If online compute resources are available, and operations can allow a several second buffer for computation, using \ourmethod~online further improves in-flight risk mitigation, in some cases without compromising distance to the target. For dynamics which online computation of \ourmethod~may take too much time, an \ourmethodoff~controller can still be used for in-flight risk mitigation.

\section{Conclusion}
In this paper, we presented \ourmethod~for UAV risk assessment and risk-bounded control using reachability analysis. We evaluate \ourmethod~in simulation and show how the approach can be used for offline and online risk mitigation. Future work includes handling multiple hazard causes at a time, integrating alternative severity models, adding objective terms that balance distance to target and risk, and extending to non-linear and continuous-time dynamics.

\bibliographystyle{IEEEtran}
\bibliography{refs}


\end{document}